\documentstyle[12pt,rsfs]{article}

\title{\Large \bf
Longitudinal wavevector- and frequency-dependent
dielectric constant of the TIP4P water model
\vspace{6pt}}

\author{{\sc Igor~P.~Omelyan} \\ \\
Institute for Condensed Matter Physics, \\
the National Ukrainian Academy of Sciences, \\
1~Svientsitsky St., UA-290011 Lviv, Ukraine\thanks
{E-mail: nep@icmp.lviv.ua} \date{}}
\newcommand{\bms}[1]{\mbox{\boldmath $#1$}}
\newcommand{\bvs}[1]{\mbox{\scriptsize\boldmath $#1$}}
\newcommand{\scs}[1]{_{\stackrel{\ }{#1}}}

\begin{document}

\setlength{\abovedisplayskip}{18pt plus4pt minus6pt}
\setlength{\belowdisplayskip}{\abovedisplayskip}
\setlength{\abovedisplayshortskip}{12pt plus2pt minus4pt}
\setlength{\belowdisplayshortskip}{\abovedisplayshortskip}

\maketitle

\vspace{24pt}

\begin{abstract}  A computer adapted theory for self-consistent
calculations of the wavevector- and frequency-dependent dielectric
constant for interaction site models of polar systems is proposed.
A longitudinal component of the dielectric constant is evaluated for
the TIP4P water model in a very wide scale of wavenumbers and
frequencies using molecular dynamics simulations. It is shown that
values for the dielectric permittivity, calculated within the exact
interaction site description, differ in a characteristic way from
those obtained by the point dipole approximation which is usually
used in computer experiment. It is also shown that the libration
oscillations, existing in the shape of longitudinal time-dependent
polarization fluctuations at small and intermediate wavevector
values, vanish however for bigger wavenumbers. A comparison between
the wavevector and frequency behaviour of the dielectric constant
for the TIP4P water and the Stockmayer model is made. The static
screening of external charges and damping of longitudinal electric
excitations in water are considered as well. A special investigation
is devoted to the time dependence of dielectric quantities in the
free motion regime.

\end{abstract}

\newpage
\section{Introduction}

\hspace{1em}  The study of dielectric properties of polar liquids by computer
experiment is still a major challenge, given that the dielectric quantities
are very sensitive to long-range intermolecular interactions and because
long trajectories are required in order to obtain adequate statistical
accuracy. For this reason, until now, the calculation of the wavevector-
and frequency-dependent dielectric permittivity, $\varepsilon(k,\omega)$,
in the whole region of $k$ and $\omega$ has been performed for the simplest
model of polar systems only, namely, for the Stockmayer fluid [1]. This model
is related to a class of molecular models describing interactions by point
dipoles embedded in molecules. The Stockmayer system, however, does not
reproduce satisfactorily the dielectric behaviour of any real polar fluid.

At the same time, more realistic interaction site (IS) models have also
been considered [2--16]. In these atomic models, intermolecular potentials
are presented as a sum of pairwise additive site-site terms. Although IS
models do not take into account internal degrees of freedom such as
electronic polarizability and, as a rule, intramolecular vibrations, they
are able to reproduce experimental results in a more satisfactory way.
There are two approaches to investigate the dielectric properties of IS
models in both computer experiment and theory. In the first, point dipole
(PD) approach, the charged sites of each molecule are replaced by a point
dipole, located in the molecular centre of mass, at constructing the reaction
field and the microscopic polarization vector. The site-site contributions
are taken into account implicitly only, namely, in the intermolecular
potentials, calculating statistical averages. In the second, IS description,
the reaction and the polarization vector are constructed in view of explicit
details of charge distribution within the molecule [16--18].

Main attention in early studies was directed to evaluate the static and
frequency-dependent dielectric constant, $\varepsilon(\omega)$, in the
long-wavelength limit. In particular, such quantities have been calculated
in molecular dynamics (MD) simulations for the rigid MCY [5] and TIP4P [6]
models of water as well as for the flexible SPC potential [7]. The frequency
dependence of the dielectric permittivity at nonzero wavevectors was
investigated for the MCY [2] and TIP4P [9, 10] water models, for models of
methyl cyanide [3, 4], methanol [11] and methanol-water mixtures [13]. But
the investigations have been restricted to small wavevector values for which
the PD approach and the IS description are expected to be identical. The
wavevector dependence, $\varepsilon(k)$, including high wavevector values,
was considered in the static limit for methanol [8], methanol-water mixtures
[12] and for the MCY water [14], using, however, the PD approximation. And
only quite a few simulations [15, 16] have been performed to define the
entire wavevector dependence within the exact IS description. To our
knowledge, there are no computer experiment data concerning the entire
frequency dependence of the dielectric permittivity of IS models at
arbitrary wavenumbers.

It is necessary to point out also the following aspect of computer simulation
of IS models. The calculation of the dielectric constant by MD requires
explicit considerations of a finite-size medium with periodic boundaries.
Usually either the cumbersome Ewald summation technique [19] or the reaction
field [20] is applied for treating long ranged interactions. As is now
well established for systems of point dipoles, proper calculations can be
made with the both methods [21]. In the case of IS models the pattern is
different. The usual point dipole reaction field (PDRF) geometry, being
exact for macroscopic systems of point dipoles, may not be necessarily
applicably to interpret simulation results of IS models [22]. Recently [16],
it has been shown by actual calculations that uncertainties of the dielectric
quantities are significant if the standard PDRF geometry is used in computer
simulations. As a result, an alternative approach has been proposed. This
approach deals with the IS description at constructing the reaction field
and leads to the so-called interaction site reaction field (ISRF) geometry.
It has been demonstrated that the ISRF geometry exhibits to be much more
efficient with respect to the usual PDRF for the investigation of dielectric
properties of IS models. Within the ISRF geometry, the longitudinal
component, $\varepsilon \scs{\rm L}(k)$, of the wavevector-dependent
dielectric constant for the MCY water model has been evaluated in a wide
wavenumber range by Monte Carlo simulations [16]. Using correct microscopic
variables for the polarization vector allows one to achieve the true infinite
wavevector behaviour $\varepsilon \scs{\rm L}(k \to \infty) = 1$.

In the present paper we propose a self-consistent theory for the calculation
of the longitudinal wavevector- and frequency-dependent dielectric constant,
$\varepsilon \scs{\rm L}(k,\omega)$, for IS models in computer experiment.
This theory is applied to the TIP4P potential using MD simulations. We show
that the PD approach is valid for describing the frequency dependence of
the dielectric constant at very small wavenumbers only. For greater
wavenumber values, the influence of higher order multipoles becomes
important. Within the IS description, the longitudinal permittivity
is evaluated in a rather very large scale of wavenumbers and frequencies
up to the infinite wavevector and frequency limit, where $\varepsilon
\scs{\rm L}(k,\omega) \to 1$.

\vspace{12pt}

\section{The reaction field for IS models}

\hspace{1em}  We consider a polar system with $N$ identical molecules
composed of $M$ interaction sites which are enclosed in a volume $V$.
The microscopic field, created by the molecules at point $\bms{r} \in
V$ and time $t$ for a macroscopic system is equal to \ $\bms{\hat {\cal
E}}(\bms{r},t) = \lim_{N \to \infty} \sum_{i=1}^N \sum_{a=1}^M \linebreak
q \scs{a} \{\bms{r}-\bms{r}_i^a(t)\}\Big/|\bms{r}-\bms{r}_i^a(t)|^3$, where
$q \scs{a}$ and $\bms{r}_i^a(t)$ are the charge and position of site $a$
within the molecule $i$, and the sum extends over all molecules and charged
sites. This sum, however, can not be calculated exactly in computer
experiment which deals with finite samples. Therefore, we must restrict
ourselves to a finite set of the terms for which $|\bms{r}-\bms{r}_i^a(t)|
\le R$, where $R$ is a cut-off radius. The radius $R$ does not exceed half
the cell length $\sqrt[3]{V}$ if a cubic sample and the toroidal boundary
conventional (tbc) are used in simulations.

Now the following problem appears. How to estimate the cut-off field caused
by the summation over the unaccessible region $|\bms{r}-\bms{r}_i^a(t)| > R$
? A solution of this problem can be found within the ISRF geometry [16]. The
result for conducting boundary conditions is
\begin{equation}
\bms{\hat {\cal E}}(\bms{r},t) = \sum_{i, a}^{N, M} q \scs{a}
\Theta (R-|\bms{r}-\bms{r}_i^a(t)|) \left(
\frac{\bms{r}-\bms{r}_i^a(t)}{|\bms{r}-\bms{r}_i^a(t)|^3}-
\frac{\bms{r}-\bms{r}_i^a(t)}{R^3} \right) \ ,
\end{equation}
where $\Theta$ is the Heviside function, i.e., $\Theta(\rho)=1$ if $\rho
\ge 0$ and $\Theta(\rho)=0$ otherwise. The first term in the right-hand side
of (1) describes the usual Coulomb field, while the second contribution
corresponds to the reaction field in the IS description. Performing the
spatial Fourier transform of (1) one obtains
\begin{equation}
\bms{\hat {\cal E}}(\bms{k},t) = \int \limits_{V,\,{\rm tbc}}
{\!\!\rm d} \bms{r}\,
{\mbox{\large e}}^{-{\rm i}\bvs{k\!\cdot\!r}} \bms{\hat {\cal E}}
(\bms{r},t) = -4\pi \bigg( 1-3 \frac{j\scs{1}(kR)}{kR}
\bigg) \bms{\hat P}_{\rm L}(\bms{k},t) \ ,
\end{equation}
where
\begin{equation}
\bms{\hat P}_{\rm L}(\bms{k},t)=
\frac{{\rm i}\bms{k}}{k^2} \sum \limits_{i, a}^{N, M}
q \scs{a} {\mbox{\large e}}^{-{\rm i} \bvs{k\!\cdot\!r}_i^a(t)}
\end{equation}
is the longitudinal component for the microscopic operator of polarization
density for IS models [18] and $j\scs{1}(z)=- \cos(z)/z+\sin(z)/z^2$ denotes
the spherical Bessel function of first order. It is worth mentioning that
as far as relativistic effects are excluded from our consideration, the
electric field (1) does not take into account retardation corrections.
For the same reason, we have neglected also the contribution to the field
caused by dynamical magnetic fields of moving charges. Because of this,
the electric field has a static form and depends on time implicitly only.
This field is pure longitudinal and can be defined uniquely via the
longitudinal component of the polarization vector $\bms{\hat P}_{\rm L}
(\bms{k},t)$, that is confirmed by equation (2).

Let us apply an external electric field $\bms{E} \scs{0}(\bms{k},t)$ to
the system under consideration, so that the total field is $\bms{\hat E}
(\bms{k},t)=\bms{E} \scs{0}(\bms{k},t)+\bms{\hat {\cal E}}(\bms{k},t)$.
The longitudinal, wavevector- and frequency-dependent dielectric constant
is defined via the material relation $4\pi \bms{P}_{\rm L}(\bms{k},\omega)=
\Big(\varepsilon \scs{\rm L}(k,\omega) - 1 \Big) \bms{E}_{\rm L}(\bms{k},
\omega)$, where $\bms{P}_{\rm L}(\bms{k},\omega)=\left< \bms{\hat P}_{\rm L}
(\bms{k},\omega) \right>$ and $\bms{E}_{\rm L}(\bms{k},\omega)=\left< \bms
{\hat k} \bms{\hat k} \bms{\cdot} \bms{\hat E}(\bms{k},\omega) \right>$ are
longitudinal components of the macroscopic polarization and total field,
$\left< \ \ \right>$ denotes the statistical averaging at the presence of
the external field, $\bms{\hat k}=\bms{k}/k$, $k=|\bms{k}|$ and the time
Fourier transform ${\scr F}(\bms{k},\omega) = \int_{-\infty}^{\infty} {\rm
d}t\, {\mbox{\large e}}^{-{\rm i} \omega t} \ {\scr F}(\bms{k},t)$ has been
used for the functions $\bms{\hat P}_{\rm L}(\bms{k},t)$ and $\bms{\hat E}
(\bms{k},t)$. The perturbation theory of first order with respect to
$\bms{E} \scs{0}$ yields $\langle \bms{\hat P}_{\rm L}(\bms{k},\omega)
\rangle=\displaystyle -\int_{0}^{\infty} {\rm d} t\, {\mbox{\large e}}^
{-{\rm i}\omega t} \frac{{\rm d}}{{\rm d} t} \left< \bms{\hat P}_{\rm L}
(\bms{k},0) \bms{\cdot} \bms{\hat P}_{\rm L}(-\bms{k},t) \right> \scs{0}
\bms{E} \scs{0}(\bms{k},\omega) \Big/ V k_{\rm B} T$, where $\left< \ \
\right> \scs{0}$ denotes the equilibrium average at the absence of the
external field, and $k_{\rm B}$ and $T$ are the Boltzmann's constant and
the temperature of the system, respectively. Then, eliminating $\bms{E}
\scs{0}(\bms{k},\omega)$ from the presented above expressions, we obtain
the desired fluctuation formula
\begin{equation}
\frac{\varepsilon \scs{\rm L}(k,\omega) - 1}
{\varepsilon \scs{\rm L}(k,\omega)}=\frac{9y
{\scr L}_{{\rm i}\omega} \Big(-\dot G_{\rm L}(k,t) \Big)}
{1+27y {\scr L}_{{\rm i}\omega} \Big(-\dot G_{\rm L}(k,t) \Big)
j\scs{1}(kR)/(kR)} = 9y {\scr L}_{{\rm i}\omega} \Big(-\dot g \scs
{\rm L}(k,t) \Big) \ .
\end{equation}
Here
\begin{equation}
G_{\rm L}(k,t)= \frac{\left<\bms{\hat P}_{\rm L}(\bms{k},0) \bms{\cdot}
\bms{\hat P}_{\rm L}(-\bms{k},t) \right> \scs{0}}{N \mu^2}
\end{equation}
is the longitudinal component of the wavevector-dependent dynamical Kirkwood
factor for the finite sample, $\mu=|\bms{\mu}_i|=|\sum_a^M q \scs{a}
\bms{r}_i^a|$ denotes the permanent magnitude of molecule's dipole moment,
${\scr L}_{{\rm i}\omega} ( ... ) = \int_{0}^{\infty} ... \ {\mbox{\large
e}}^{-{\rm i} \omega t} {\rm d} t$ designates the Laplace transform,
$y=4\pi N \mu^2 \Big/ 9Vk_{\rm B}T$ and $\dot G_{\rm L}(k,t) \equiv \partial
G_{\rm L}(k,t)/\partial t$. In the case $R \to \infty$, when $j\scs{1}(kR)
/(kR) \to 0$, the computer adapted formula (4) reduces to the fluctuation
formula for macroscopic systems in terms of the infinite-system Kirkwood
factor $g \scs{\rm L}(k,t)=\lim_{R \to \infty} G_{\rm L}(k,t)$.

It is essential to emphasize that the fluctuation formula (4) takes into
account finiteness of the system explicitly by the factor $j\scs{1}(kR)/
(kR)$. As a consequence, for sufficiently large systems, the bulk ($N, V
\to \infty$) dielectric constant can be reproduced via the finite-system
Kirkwood factor $G_{\rm L}(k,t)$ calculated in simulations. However, in
order to achieve this self-consistency in the evaluation of the bulk
dielectric constant, the equilibrium averaging in $G_{\rm L}(k,t)$
must be performed for systems with the intermolecular potential which
leads exactly to the microscopic electric field $\bms{\hat {\cal E}}
(\bms{r},t)$ (1) at which the fluctuation formula has been derived.
This intermolecular potential can be determined via the relation $\bms
{\hat {\cal E}}(\bms{r},t)=-\bms{\nabla} {\mit \Phi}(\bms{r},t)$), whence
${\mit \Phi}(\bms{r},t)=\sum_{i, a}^{N, M} \phi_i^a(\bms{r},t)$, where
$\phi_i^a(\bms{r},t) = q \scs{a} \{1/\rho_i^a(t) + \frac12 {\rho_i^a(t)}^2
/R^3+C\}$, $\rho_i^a(t)=|\bms{r}-\bms{r}_i^a(t)|$ and $C$ is, in general,
an arbitrary constant which for infinite systems is chosen as $\phi_i^a |_
{\rho_i^a \to \infty}=0$. In our case, according to the toroidal boundary
conventional, $\phi_i^a |_{\rho_i^a = R}=0$ and, therefore, $C=-3/2 \,
R^{-1}$. Then the intermolecular potential of interaction is $\sum_{a,b}^M
q \scs{b} \phi_i^a(\bms{r}_j^b(t)) = \sum_{a,b}^M q \scs{a} \phi_j^b
(\bms{r}_i^a(t))$, i.e,
\begin{equation}
\varphi \scs{ij}^{\rm IS} = \sum_{a,b}^M q \scs{a} q \scs{b}
\Theta (R-|\bms{r}_i^a-\bms{r}_j^b|) \left( \frac{1}{|\bms{r}_i^a-
\bms{r}_j^b|} + \frac12 \frac{|\bms{r}_i^a-\bms{r}_j^b|^2}{R^3} - \frac{3}
{2R} \right) \ ,
\end{equation}
where the site-site cut-off $|\bms{r}_i^a-\bms{r}_j^b| \le R$ is performed.

The fluctuation formula (4) has already been applied previously in the
long-wavelength limit at the investigation of the static and
frequency-dependent dielectric constant for the MCY [5] and TIP4P [6]
models. However, acting within a semiphenomenological framework, it
was not understood how to perform the truncation for intermolecular
potentials. As a result, the molecular cut-off $r_{ij}=|\bms{r}_i-\bms{r}_j|
\le R$, where $\bms{r}_i$ is the centre of mass of the $i$th molecule, and
the usual PDRF have been suggested:
\begin{equation}
\varphi \scs{ij}^{\rm PD} = \Theta (R-r_{ij}) \left( \sum_{a,b}^M
\frac{q \scs{a} q \scs{b}}{|\bms{r}_i^a-\bms{r}_j^b|} -
\frac{\bms{\mu}_i \bms{\cdot} \bms{\mu}_j}{R^3} \right) \ .
\end{equation}
It is easy to show that the self-consistent potential (6) can be reduced
to (7) in one case only, namely, when $d/R \to 0$, where $d=2 \max_a
|\bms{r}_i^a-\bms{r}_i|$ is the diameter of the molecule. In this case,
the positions for sites and the centre of mass are undistinguished within
the same molecule. For IS finite samples, where $d/R \ne 0$, the PDRF
potential (7) may affect on a true macroscopic behaviour of the system
considerably. At the same time, the intermolecular potential $\varphi
\scs{ij}^{\rm IS}$ corresponds completely to the conditions at which the
fluctuation formula (4) has been obtained. Therefore, as far as this
formula is applied to treat simulation results, the ISRF potential (6),
instead of (7), must be used in computer experiment to reproduce a correct
value for the dielectric constant.

Nevertheless, the molecular cut-off scheme can also be acceptable, but
the PDRF potential (7) as well as the fluctuation formula (4) need to be
modified. In fact, the intermolecular potential $\varphi \scs{ij}^{\rm PD}$
takes into account dipole contributions only into the reaction field.
Additional terms can be identified within the molecular reaction field
(MRF) approach. The result is [16]:
\begin{equation}
\varphi \scs{ij}^{\rm M} = \Theta (R_d-r_{ij}) \left(
\sum_{a,b}^M \frac{q \scs{a} q \scs{b}} {|\bms{r}_i^a-
\bms{r}_j^b|} - \frac{\bms{\mu}_i \bms{\cdot} \bms{\mu}_j}{R^3} -
\frac {\bms{\rm q}_i \bms{:} \bms{\rm q}_j - 3 (\bms{\rm q}_i \bms{:}
\bms{\mu}_j \bms{r}_{ij} + \bms{\rm q}_j \bms{:} \bms{\mu}_i \bms{r}_{ji})}
{6R^5} + \ldots \right) ,
\end{equation}
where $\bms{\rm q}_i=\sum_a^M q \scs{a} (3 \bms{\delta}_i^a \bms{\delta}_i^a
-{{\delta}_i^a}^2 \bms{\rm I})$ is the quadrupole moment of the $i$th
molecule with respect to its centre of mass, $\bms{\delta}_i^a=\bms{r}_i^a-
\bms{r}_i$, $\bms {\rm I}$ is the unit tensor, $R_d =R-d/2$, and multipoles
of higher orders have been neglected. The fluctuation formula corresponding
to the potential (8) can be derived as follows.

The microscopic electric field created by the molecules at point $\bms{r}$,
coinciding with the centre of the spherical cavity of radius $R$, surrounded
by the infinity conduction medium, can be presented in the MRF geometry as
\begin{equation}
\bms{\hat {\cal E}}(\bms{r},t)=\sum_{i=1}^N
\Theta (R_d-|\bms{r}-\bms{r}_i(t)|)
\left\{ \sum_{a=1}^M q \scs{a} \frac{\bms{r}-\bms{r}_i^a(t)}
{|\bms{r}-\bms{r}_i^a(t)|^3} + \frac{\bms{\mu}_i(t)}{R^3} \right\}
\end{equation}
(quadrupole and higher-order moments do not contribute the field in the
centre of the cavity). Performing the spatial Fourier transform of (9)
yields after some algebra
\begin{equation}
\bms{\hat {\cal E}}(\bms{k},t) = -4\pi \left\{ \sum_{\bvs{k'}}
\bms{\hat P}(\bms{k},\bms{k'},t) \frac{3 j\scs{1} (|\bms{k}-\bms{k'}|R_d)}
{|\bms{k}-\bms{k'}|R_d} - \frac{j\scs{1}(kR_d)}{kR_d} \frac{R_d^3}{R^3}
\bms{\hat M}(\bms{k},t) \right\} \ ,
\end{equation}
where
\begin{equation}
\bms{\hat P}(\bms{k},\bms{k'},t)=\frac{{\rm i}\bms{k'}}{k'^2} \sum_{i=1}^N
{\mbox{\large e}}^{-{\rm i} \bvs{k\!\cdot\!r}_i(t)} \sum_{a=1}^M q \scs{a}
{\mbox{\large e}}^{-{\rm i} \bvs{k'\!\cdot\!\delta}_i^a(t)} \ , \ \ \ \ \
\bms{\hat M}(\bms{k},t)=\sum_{i=1}^N \bms{\mu}_i(t)
{\mbox{\large e}}^{-{\rm i} \bvs{k\!\cdot\!r}_i(t)} \ ,
\end{equation}
$\bms{k}, \bms{k'}$ are one of the allowed wavevectors $\bms{n} k_{\rm
min}$ of the reciprocal lattice, $\bms{n}$ designates a vector with integer
components and $k_{\rm min}=2 \pi/\sqrt[3]{V}$. As we can see $\bms{\hat P}
(\bms{k},\bms{k},t) \equiv \bms{\hat P}_{\rm L}(\bms{k},t)$, while $\bms{\hat
M}(\bms{k})$ is the microscopic operator of polarization density for a point
dipole system. Its longitudinal component $\bms{\hat M}_{\rm L}(\bms{k},t)=
\bms{\hat k} \bms{\hat k} \bms{\cdot} \bms{\hat M}(\bms{k},t)$ coincides
with $\bms{\hat P}_{\rm L}(\bms{k},t)$ in the point dipole limit: $d \to 0$,
$q \scs{a} \to \infty$, provided $\mu \to $ const. Using the internal field
(10) and applying the perturbation theory with respect to the external field,
similar to that as in the case of the ISRF geometry, we obtain the following
fluctuation formula
\begin{equation}
\frac{\varepsilon \scs{\rm L}(k,\omega) \!-\! 1}
{\varepsilon \scs{\rm L}(k,\omega)}=
\frac{9y {\scr L}_{{\rm i}\omega}(-\dot G_
{\rm L}(k,t))}{1\!-\!27y \Bigg\{ \displaystyle
\frac{{\scr L}_{{\rm i}\omega}(\dot{Q}
\scs{\rm L}(k,t)) j \scs{1}(kR_d)}{3kR_d}
\frac{R_d^3}{R^3} - \sum_{\bvs{q} (q \ne 0)} \!\!
\frac{j\scs{1}(q R_d)}{q R_d}
{\scr L}_{{\rm i}\omega}(\dot G_
{\rm L}(\bms{k},\bms{q},t)) \Bigg\}} \ ,
\end{equation}
where the summation is extended over the infinite set $\bms{n} k_{\rm min}$
of nonzero wavevectors $\bms{q}$, $G_{\rm L}(\bms{k},\bms{q},t)\!=\!
\left< \bms{\hat P}_{\rm L}(\bms{k},\bms{k}\!+\!\bms{q},0) \bms{\cdot}
\bms{\hat P}_{\rm L}(-\bms{k},t) \right> \scs{0} \!\Big/\! N \mu^2$, \
$Q \scs{\rm L}(k,t)\!=\!\left< \bms{\hat M}_{\rm L}(\bms{k},0) \bms{\cdot}
\bms{\hat P}_{\rm L}(-\bms{k},t) \right> \scs{0} \!\Big/\! N \mu^2$ are
time correlation functions and $\bms{\hat P}_{\rm L}(\bms{k},\bms{k'})=
\bms{\hat k} \bms{\hat k} \bms{\cdot} \bms{\hat P}(\bms{k},\bms{k'})$.

As was mentioned earlier, internal electric fields of classical systems
are pure longitudinal. However, as it follows from the structure of
equations (11), the field $\bms{\hat {\cal E}}(\bms{k},t)$ (10) for finite
systems may contain a transverse component as well which vanishes in the
limit $R \to \infty$ of infinite systems only. At the same time, in the ISRF
geometry the field $\bms{\hat {\cal E}}(\bms{k},t)$ (2) remains by a
longitudinal one even for finite systems. Therefore, from a physical
point of view, the ISRF geometry is more natural than the MRF approach,
because it does not influence on the true structure of electric fields.
Moreover, within the MRF geometry, the fluctuation formula (12) appears to
be very complicated with respect to the much more simple formula (4) in the
ISRF approach. It requires the knowledge of the additional correlation
functions $Q \scs{\rm L}(k,t)$ and $G_{\rm L}(\bms{k},\bms{q},t)$. While
the function $Q \scs{\rm L}(k,t)$ can be evaluated in the usual way, the
function $G_{\rm L}(\bms{k},\bms{q},t)$ at fixed $t$, even for spatially
homogeneous systems, depends upon three parameters (magnitudes of $\bms{k}$,
$\bms{q}$ and the cosine between them) and its calculation is a hard problem
and impractical in simulations. Finally, the site-site cut-off truncation (6)
has yet a minor advantage over the molecular cut-off scheme (8) because the
intermolecular potential of interaction $\varphi \scs{ij}^{\rm IS}$ is
continuous and continuously differentiable. This avoids the system energy
drift associated with the passage of sites through the surface of the
truncation sphere.

\vspace{12pt}

\section{Application to the TIP4P potential}

\hspace{1em}  MD simulations have been carried out for the TIP4P potential
[23] in the microcanonical ensemble at a density of 1 g/cm$^3$ and at a
temperature of $T=293$ K using the ISRF geometry (6). We considered $N=256$
molecules in the cubic volume $V=L^3$ to which the toroidal boundary
conditions have been applied. The interaction cut-off radius was half the
cell length, $R=L/2=9.856$\AA. The simulations were started from a well
equilibrated configuration for the positions of sites, obtained by Monte
Carlo simulations. Initial velocities of the molecules were generated at
random with the Maxwell distribution. The equations of motion were integrated
with a time step $\Delta t=2$ fs on the basis of a matrix method using the
Verlet algorithm in velocity form. The system was allowed to achieve
equilibrium for 100 000 time steps. The observation time over the equilibrium
state was 1 000 000 $\Delta t=2$ ns and each 10th time step has been
considered to compute the equilibrium averages. In order to provide the
exact conservation for the total energy of the system during such a rather
very long observation time, the velocities of atoms were slightly rescaled
after every 500 time steps, so that the relative total energy fluctuations
did not exceed 0.01\% over the whole run. The time correlation functions
were calculated with the time step $\Delta t$ in the interval of $1000
\Delta t=2$ ps. The wavevector-dependent quantities were investigated for
$k=[0,1,\ldots,300] k_{\rm min}$, where $k_{\rm min}=2 \pi/L=0.319{\rm
\AA}^{-1}$.

The longitudinal component $G_{\rm L}(k) \equiv G_{\rm L}(k,0)$ of the
wavevector-dependent Kirkwood factor (5) at $t=0$, obtained in the
simulations within the IS description, and the corresponding function
$g \scs{\rm L}(k) \equiv g \scs{\rm L}(k,0)$ (4), related to the infinite
system, are shown in fig.~1 by dashed and solid curves, respectively. In
the PD approximation the diameter $d$ of the molecule is assumed to be
sufficiently small in order to be entitled to replace the true microscopic
operator $\bms{\hat P}_{\rm L}(\bms{k},t)$ (3) of polarization density for
an IS system by its analogue $\bms{\hat M}_{\rm L}(\bms{k},t)=\bms{\hat k}
\sum_{i=1}^N \bms{\hat k} \bms{\cdot} \bms{\mu}_i(t) {\mbox{\large e}}^
{-{\rm i} \bvs{k\!\cdot\!r}_i(t)}$ for a system of point dipoles. The
infinite-system Kirkwood factor calculated within the PD approximation,
$\left<\bms{\hat M}_{\rm L}(\bms{k},0) \bms{\cdot} \bms{\hat M}_{\rm L}
(-\bms{k},0) \right> {_{\!\!}} \scs{0}\Big/N \mu^2$, is presented in fig.~1
by open squares. For the TIP4P molecule $d=1.837$\AA\ and, therefore, the
PD approximation can not be used for calculating the wavevector-dependent
dielectric quantities at $k \mathop{_\sim} \limits^{{\mbox{\footnotesize
$>$}}} 2 \pi/d \sim 3.4{\rm \AA}^{-1}$. Indeed, as we can see from the
figure, the PD approach reproduces values for the Kirkwood factor
satisfactorily in a small region of wavenumbers only, namely, at $k < 2.5
{\rm \AA}^{-1}$. At greater wavevectors, these values differ considerably
from those obtained within the IS description. In the infinite wavevector
regime ($k \to \infty$) the Kirkwood factor tends to $1/3$, when the PD
approximation is used, whereas it approaches zero with the asymptotic
behaviour $g^{\rm (f)} \scs{\rm L}(k)$ within the IS description (see
Appendix, equation (A4)). The function $g^{\rm (f)} \scs{\rm L}(k)$ is
plotted in fig.~1 by the long-short dashed curve.

In order to verify explicitly that the ISRF geometry reproduces adequately
the wave\-vector dependence of dielectric quantities, we have performed
also additional simulations for calculations of $g \scs{\rm L}(k)$ using
the generally recognized Ewald geometry and the PDRF potential (7). These
results are included in fig.~1 as well (the open circles and dotted curve).
The parameters $\eta =5.76/L$ and $k_{\rm max} = 5 k_{\rm min}$ were used
in the Ewald summation of Coulomb forces. As we can see from the figure,
the ISRF method leads to results which are identical to those obtained
within the cumbersome Ewald technique. At the same time, deviations of
values for the wavevector-dependent Kirkwood factor obtained using the
PDRF potential from those evaluated within the ISRF geometry are of order
$20\%$. They are well exhibited at intermediate values of wavevectors.
Such a situation can be explained by the fact that the PDRF geometry does
not take into account the spatial distribution of charges within the
molecule at constructing the reaction field and, thus, the precision
of the calculations for wavevector-dependent dielectric quantities at
$k \sim 2 \pi/d$ can not exceed $d/R \sim 20\%$. And only for great
wavevector values $(k>6{\rm \AA}^{-1})$, where the influence of boundary
conditions is negligible, the both geometries become equivalent between
themselves and the analytical formula (A4) can be applied here.

To analyze the wavevector-dependent dielectric function in the low frequency
limit, $\omega \to 0$, it is more convenient to rewrite the fluctuation
formula (4) in terms of the infinite-system Kirkwood factor as follows
\begin{equation}
\varepsilon \scs{\rm L}(k,\omega) = \frac{(1-9y g \scs{\rm L}(k))-
9y {\rm i}\omega {\scr L}_{{\rm i}\omega} (g \scs{\rm L}(k,t))}
{(1-9y g \scs{\rm L}(k))^2 + (9y)^2 \omega^2 {\scr L}_{{\rm i}\omega}^2
(g \scs{\rm L}(k,t))} \ ,
\end{equation}
where the equality ${\scr L}_{{\rm i}\omega} (-\dot g \scs{\rm L}(k,t))=g
\scs{\rm L}(k) -{\rm i}\omega {\scr L}_{{\rm i}\omega} (g \scs{\rm L}(k,t))$
has been used. In particular, the static wavevector-dependent dielectric
function $\varepsilon \scs{\rm L}(k) = 1 \Big/(1-9y g \scs{\rm L}(k))$ is
obtained from the relation (13) putting $\omega=0$. This function, calculated
within the IS description and the PD approximation, is displayed in fig.~2a
as circles connected by the solid curve and as open squares, respectively.
As has been pointed out previously [6], the static dielectric constant of
the TIP4P water at infinite wavelengths ($k \to 0$) appears to be smaller
than the experimental value $\varepsilon\scs{0} \approx 80$ for real water
and consists about 53. This value has been determined on the basis of MD
simulations using the PDRF geometry. In our calculations within the exact
ISRF geometry we have found that $\varepsilon \scs{0} \equiv \varepsilon
\scs{\rm L}(k \to 0) \approx 50$. Therefore, the TIP4P water reproduces the
static dielectric constant of real water even somewhat worse than this has
been established earlier. The function $\varepsilon \scs{\rm L}(k)$ in the
PD approximation behaves like that for the Stockmayer fluid [21, 24]. For
example, in the infinite wavevector limit, the dielectric constant tends
to the wrong Onsager limit value $1/(1-3y)=-0.0649$.

The true wavevector behaviour for the dielectric constant can be reproduced
within the IS description only. As we can see, the static dielectric constant
of the TIP4P water is negative over a wavevector range bounded by two
singularities, where $9y g \scs{\rm L}(k) \to 1$ and $\varepsilon \scs{\rm
L}(k) \to \pm \infty$. The first singularity is achieved at $k=k \scs{\rm I}
\approx 0.297{\rm \AA}^{-1}$, whereas the second singularity may be estimated
[14] via the relation
\begin{equation}
k \scs{\rm II} \approx \sqrt{\frac{9y}{\mu^2} \sum_{a=1}^M q \scs{a}^2} =
\sqrt{\frac{4\pi N}{V k_{\rm B}T}\sum_{a=1}^M q \scs{a}^2} \ .
\end{equation}
This relation is derived solving the equation $9y g^{\rm (f)} \scs{\rm L}
(k)=1$ and neglecting intersite contributions (the second term in the
right-hand side of (A4)) in the intramolecular part $g^{\rm (f)} \scs{\rm L}
(k)$ of the Kirkwood factor. Applying the relation (14) yields for the TIP4P
molecule $k \scs{\rm II} \approx 19.71{\rm \AA}^{-1}$, that is in an
excellent accord with the simulation result $20.28{\rm \AA}^{-1}$. At $k>k
\scs{\rm II}$ the function $\varepsilon \scs{\rm L}(k)$, remaining positive,
slightly approaches unity at $k \mathop{_\sim} \limits^{{\mbox{\footnotesize
$>$}}} 100{\rm \AA}^{-1}$. In the low frequency limit at $k^\star=k \scs{\rm
I}, k \scs{\rm II}$ we put $1-9y g \scs{\rm L}(k^\star)=0$ and obtain from
(13) that the dielectric constant behaves as $\varepsilon \scs{\rm L}
(k^\star,\omega)=-{\rm i}\Big/ \omega \tau_{\rm L}^{\rm cor}(k^\star)$, where
$\tau_{\rm L}^{\rm cor}(k)=\int_0^\infty g \scs{\rm L}(k,t)/g \scs{\rm L}(k)
{\rm d} t$ is the correlation time. Therefore, the real part of the
dielectric permittivity in this regime is equal to zero, whereas the
imaginary part tends to infinity as $1/\omega$. This result can be better
understood introducing the generalized coefficient $\sigma \scs{\rm L}(k,
\omega)$ of polarization conductivity, which defines the macroscopic
current $\bms{I}_{\rm L}(\bms{k},t)=\partial \bms{P}_{\rm L}(\bms{k},t)/
\partial t$ in the frequency representation as $\bms{I}_{\rm L}(\bms{k},
\omega)=\sigma \scs{\rm L}(k,\omega) \bms{E}_{\rm L}(\bms{k},\omega)$.
Taking into account that in this representation $\bms{I}_{\rm L}
(\bms{k},\omega)={\rm i} \omega \bms{P}_{\rm L}(\bms{k},\omega)$, we
can express the polarization conductivity in terms of the dielectric
constant as $\sigma \scs{\rm L}(k,\omega)=\frac{{\rm i} \omega}{4\pi}
(\varepsilon \scs{\rm L}(k,\omega)-1)$. Thus in the limit of small
frequencies we have that $\sigma \scs{\rm L}(k)=\lim \limits_{\omega
\to 0} \sigma \scs{\rm L}(k,\omega)=1\Big/ 4\pi \tau_{\rm L}^{\rm cor}
(k) \ne 0$ if $k \in k^\star$ and $\sigma \scs{\rm L}(k)=0$ otherwise.

In view of the existence of a nonvanishing coefficient for the generalized
conductivity in the static limit, the following question arises. Does it
violate the well-known law that static macroscopic currents are absent in
the dielectrics? We can be sure that this law remains in force, because
$\sigma \scs{\rm L}(k) \ne 0$ when $|\varepsilon \scs{\rm L}(k)| \to
\infty$ and, therefore, the total electric field in the system vanishes
because of $\bms{E}_{\rm L}(\bms{k})=\bms{\hat k} \bms{\hat k} \bms{\cdot}
\bms{E} \scs{0}(\bms{k}) \Big/ \varepsilon \scs{\rm L}(k) \to 0$, so that
the polarization current does not appear. In this case, the longitudinal
external field is compensated completely by the internal field of polar
molecules. Moreover, the singularities in the dielectric permittivity do
not lead to singularities in physically observing quantities. This is
so because these quantities are expressed through the external electric
field using multipliers of type $1/\varepsilon \scs{\rm L}(k)$ or
$(\varepsilon \scs{\rm L}(k)-1)/\varepsilon \scs{\rm L}(k)$ which are
free of singularities. For instance the ratio $\bms{{\cal E}}(\bms{k})
\Big/(\bms{\hat k} \bms{\hat k} \bms{\cdot} \bms{E} \scs{0}(\bms{k}))$ of
the macroscopic static field of polar molecules and the external field is
determined by the factor $-(\varepsilon \scs{\rm L}(k)-1)/\varepsilon
\scs{\rm L}(k)=-9y g \scs{\rm L}(k)$. The sign minus shows that the field
of molecules is opposite to the external field, because molecular dipole
moments align always along external fields. The maximum magnitude of this
factor ranges up to about 44.9 at $k \approx 3.03{\rm \AA}^{-1}$ (the
first maximum of the Kirkwood factor), where the dielectric constant takes
its almost maximum value $-0.0228$ within the negative region. The value
$3.03{\rm \AA}^{-1}$ corresponds to the wavelength $\lambda = 2\pi/k
\approx 2.07{\rm \AA} \sim d$. Therefore, optimal conditions (maximal
average torques acting on molecules due to electric fields) for alignment
of dipoles along external fields are observed when spatial inhomogeneity
of these fields has a characteristic length of its varying in coordinate
space of order of the molecular diameter (the so called spatial resonance).
A similar wavevector behaviour of $\varepsilon \scs{\rm L}$ is exhibited
for a central force model of water [15].

The knowledge of the dielectric constant in the whole region of wavenumbers
allows one to solve the problem of static screening of external charges
in water. An external charge, enclosed in the dielectric, causes a
polarization of the system. The potential of the electric field, created
in such a way, can be written in the $\bms{k}$-representation as
$\varphi(\bms{k})=\varphi^{(0)}(\bms{k})/\varepsilon \scs{\rm L}(k)$,
where $\varphi^{(0)}(\bms{k})=4\pi/k^2$ is the Fourier transform for the
potential $\varphi^{(0)}(\bms{r})=1/r$ of an unit charge in the vacuum. Then
applying the inverse Fourier transform to the function $\varphi(\bms{k})$
one obtains $\varphi(\bms{r}) = \frac{1}{(2\pi)^3} \int \varphi(\bms{k})
{\mbox{\large e}}^{{\rm i}\bvs{k\!\cdot\!r}} {\rm d} \bms{k}=\frac{2}{\pi}
\int \limits_0^\infty \frac{1}{\varepsilon \scs{\rm L}(k)} \frac{\sin{kr}}
{kr} {\rm d} k = \varphi^{(0)}(\bms{r}) \Big/{\varepsilon \scs{\rm L}(r)}$.
The function $1/\varepsilon \scs{\rm L}(r)$, which is plotted in fig.~2b,
describes the static screening. It can be shown easily, using the asymptotic
behaviour (A4), that the dielectric constant in the infinite wavenumber
regime is expanded over inverse wavevectors as $\varepsilon \scs{\rm L}
(k)=1+1/(r \scs{\rm D}k)^2+{\cal O}(k^{-4})$, where $r \scs{\rm D}=1/k
\scs{\rm II}$. The expression $\varepsilon \scs{\rm L}(k)=1+1/(r \scs{\rm
D}k)^2$ is often used in the whole range of wavevectors, considering
processes of static screening in plasma. This leads to the well-known
Debye exponential screening $1/\varepsilon \scs{\rm L}(r)=\exp(-r/r
\scs{\rm D})$ with the Debye's radius $r \scs{\rm D}$. In the case of
dielectrics the pattern is different. The function $1/\varepsilon \scs
{\rm L}(r)$, starting from 1 at $r=0$, exhibits a pronounced oscillatory
feature, reflecting the influence of the microscopic structure of the
system. And only beginning from distances to the external charge of order
$r \mathop{_\sim} \limits^{{\mbox{\footnotesize $>$}}} 15 {\rm \AA}$ this
function tends to its value $1/{\varepsilon \scs{0}}$ in the macroscopic
limit.

Examples for the normalized autocorrelation functions $\Phi \scs{\rm L}
(k,t) = G_{\rm L}(k,t)/G_{\rm L}(k)$ (5) and $\phi \scs{\rm L}(k,t)
= g \scs{\rm L}(k,t)/g \scs{\rm L}(k)$ (4), describing the time decay of
longitudinal polarization fluctuations in the finite ($N=256$) and infinite
systems are plotted in fig.~3 by the circles and solid curves, respectively
(the infinite-system functions are obtained applying the inverse Laplace
transform to the relation (4)). The curves are for a set of wavevectors
accessible in the simulations. For the purpose of comparison, analogous
functions, $\langle \bms{\hat M}_{\rm L}(\bms{k},0) \bms{\cdot} \bms{\hat
M}_{\rm L}(-\bms{k},t) \rangle \scs{0} \Big/ \langle \bms{\hat M}_{\rm L}
(\bms{k},0) \bms{\cdot} \bms{\hat M}_{\rm L}(-\bms{k},0) \rangle \scs{0}$,
obtained within the PD approximation for the finite system, are also
presented in this figure by dashed curves. The self ($i=j$ in equation (A1))
part $\phi^{\rm (s)} \scs{\rm L}(k,t)=g^{\rm (s)} \scs{\rm L}(k,t)/
g^{\rm (s)} \scs{\rm L}(k)$ of $\phi \scs{\rm L}(k,t)$ and the normalized
function $\phi^{\rm (f)} \scs{\rm L}(k,t)=g^{\rm (f)} \scs{\rm L}(k,t)/
g^{\rm (f)} \scs{\rm L}(k)$ (note that $g^{\rm (s)} \scs{\rm L}(k)=
g^{\rm (f)} \scs{\rm L}(k)$ in the static limit) corresponding to a
noninteracting system, calculated with the help of stochastic simulations
according to equation (A2), are included in fig.~3 as well and shown by
the dotted and long-short dashed curves, respectively. As we can see, the
self portion of $g \scs{\rm L}(k,t)$ begins to dominate already at $k \ge
10 k_{\rm min}$ and the free motion regime starts at $k \ge 40 k_{\rm min}$.
The normalized function $g^{\rm (f)} \scs{\rm L}(k,t)$, evaluated via the
analytical formula (A3), is plotted in insets (e)-(h) of fig.~3 by open
circles.

The time correlation functions obtained in the IS description are identical
to those within the PD approximation in the zero wavevector limit only but
they differ from one another significantly at greater ($k> k_{\rm min}$)
wavevector values. For example, in the infinite wavevector limit (see
fig.~3g,h), the PD functions exhibit the pure Gaussian behaviour
$f \scs{\rm G}(k,t)$, whereas the IS functions have a more complicated
structure of the time dependence (A3). Taking into account the remarks
concerning the behaviour of these functions in the static limit (see
fig.~1), we conclude that the PD approximation, being exact for systems
of point dipoles, is unsuitable, in general, to investigate the wavevector-
and frequency-dependent dielectric constant of IS models of polar fluids.

The most striking feature in the time behaviour for the longitudinal
component of the infinite-system function $\phi \scs{\rm L}(k,t)$ is damped
oscillations superimposed on the exponential. They are called librational
oscillations and can be observed for the self part $\phi^{\rm (s)} \scs{\rm
L}(k,t)$ as well. To show these oscillations in more detail, we have also
presented 3D plot of the function $\phi \scs{\rm L}(k,t)$ in fig.~4. The
librational oscillations appear for systems with a slowly relaxing character
of intermolecular torques and describe the rapid motion of a molecule in
the average electric field $\sim \frac{\bvs{r}}{r^3 \varepsilon \scs{\rm L}
(r)}$ of its neighbours (see fig.~2b). It is not a surprise, the librations
are absent in the shape of the free-motion functions $\phi^{\rm (f)}
\scs{\rm L}(k,t)$ which can be related to polarization fluctuations of a
dilute polar fluid. The librational oscillations have been found previously
for the MCY [2, 5] and TIP4P potentials [6] and for a model of methyl
cyanide [3]. However, the investigation of the librational oscillations
was restricted to zero and very small values ($\sim k_{\rm min}$) of
wavevector. In our study, performed in a rather very large scale of
wavenumbers, we have identified the librational oscillations at small and
intermediate wavenumber values, namely, at $k \mathop{_\sim} \limits^
{{\mbox{\footnotesize $<$}}} 10 k_{\rm min} \sim 3{\rm \AA}^{-1}$ (see
figs.~3, 4). They vanish at bigger wavevectors when the time correlation
functions behave like those for a noninteracting system.

It is necessary to underline that all the functions after a sufficiently
long period begin to decay purely exponentially in time as \ $\sim \exp(-t/
\tau_{\rm L}^{\rm rel}(k))$, where $\tau_{\rm L}^{\rm rel}(k)$ is the
relaxation time. The wavevector-dependent relaxation $\tau_{\rm L}^{\rm
rel}(k)$ as well as the correlation $\tau_{\rm L}^{\rm cor}(k)$ times for
the finite and infinite systems are shown in fig.~5. At $k=0$ we have found
for the finite-system function $\tau^{\rm rel}(0)=6.7$ps and $\tau^{\rm
cor}(0)=6.4$ps. As was pointed out above, the correlation functions have
been evaluated in the finite time interval $t \in$ [0, 2ps]. The
relaxation times will be used by us to calculate contributions at time
integration for $t > 2$ps, where all the correlation functions exhibit
almost purely relaxation behaviour. We indicate the longest tails in the
time correlation functions for the infinite system at $k \sim 1.7$ and
$2.9{\rm \AA}^{-1}$. These functions remain positive anywhere in time
space, contrary to the Stockmayer fluid with the dipolaron behaviour
[1, 25] of dipole moment fluctuations.

Now, we are in a position to discuss the result of calculations (4) for
the longitudinal wavevector- and frequency-dependent dielectric constant
$\varepsilon \scs{\rm L}(k,\omega)$ of the TIP4P water. Real $\varepsilon'
\scs{\rm L}(k,\omega)$ and imaginary $\varepsilon''\scs{\rm L}(k,\omega)$
parts of $\varepsilon \scs{\rm L}(k,\omega)=\varepsilon'\scs{\rm L}(k,\omega)
-{\rm i} \varepsilon''\scs{\rm L}(k,\omega)$ are shown in figs.~6 and 7
as solid and dashed curves, respectively. The calculation of the
frequency-dependent dielectric constant for the TIP4P water has already
been dealt with in the paper [6]. In this paper, however, the dielectric
relaxation was investigated at zero wavevector value only. Moreover, the
simulation results here were actually deduced within the PDRF potential
rather than the exact ISRF geometry. As has been concluded previously [16],
investigating the wavevector dependence of the dielectric constant of the
MCY model, and shown above for the wavevector-dependent Kirkwood factor
$g \scs{\rm L}(k)$ of the TIP4P water (see fig.1), it is risky to use the
PDRF geometry for systems of hundreds molecules. Although it is hard to
tell except by numerical computations whether the results [6] for the
frequency dependence of the dielectric constant differ significantly from
the exact results corresponding to the macroscopic system. For this reason
we have repeated the calculation of the frequency-dependent dielectric
constant $\varepsilon(\omega) = \lim_{k \to 0} \varepsilon \scs{\rm L}
(k,\omega)$. The result of these calculations, performed in the ISRF
geometry, is shown in fig.~6a. Comparing the previous result (cf. fig.~6
of ref.~6) with our one, we can observe that apart from the slight
differences in the low frequency regime the agreement is quite good for
higher frequencies. Therefore, the dielectric constant is less sensitive
to boundary effects at intermediate and high values of frequency.

Since all the correlation functions decay exponentially at great times, in
the low frequency regime the dielectric permittivity behaves like the Debye
dielectric for arbitrary wavevectors. Deviations can be visible at higher
frequencies and at small and intermediate wavevector values, where the
collective molecular librations take an important role in forming the
polarization fluctuations. Above some $\omega$=1-10 THz, where THz=ps$^{-1}$,
the relaxation process gradually changes into a resonance process
characterized by an frequency of 60-200 THz (which depends on wavevector)
and reflecting the rapid librational motion of the molecules. The main
differences at zero wavevector between the frequency-dependent dielectric
constant for the TIP4P water, Debye and Stockmayer models as well as real
water have been done already [6, 26]. Now we consider differences in the
behaviour on frequency between the longitudinal dielectric constant of
the TIP4P water and the Stockmayer model [1] at nonzero wavevector values.

Despite a similar overall shape of $\varepsilon \scs{\rm L}(k,\omega)$ to
that of the Stockmayer fluid, additional features emerge for the TIP4P water,
namely, a second maximum in the imaginary part $\varepsilon''\scs{\rm L}(k,
\omega)$ at $k < 7 k_{\rm min}$ and a more complicated structure for the
real part $\varepsilon'\scs{\rm L}(k,\omega)$ at $k < 20 k_{\rm min}$.
For example, the real part $\varepsilon'\scs{\rm L}(k,\omega)$ for the
Stockmayer system always increases at increasing frequency, while we can
observe clearly additional maximum and minimum for the TIP4P water in the
librational resonance range. Although the TIP4P model does not reproduce
the dielectric properties of water at all well, since it does not account
for polarizability and the coupling of inter- and intramolecular motions,
a similar structure in the dielectric constant of real water for nonzero
wavevector values may be present as well, because in real water the
librational frequency is well separated from the lowest intramolecular
vibration [27]. It is interesting also to point out that the dielectric
constant of the TIP4P water at frequencies $\omega > 10$ THz practically
does not depend on wavevector in the range $0 \le k \le 2 k_{\rm min}$
and even up to $k=10 k_{\rm min}$ at $\omega > 200$ THz and remains the
same as in the low wavevector limit (cf. fig.~6b-e). The wavevector- and
frequency-dependent dielectric constant at great wavenumber values is shown
in fig.~7. As we can see, the correct value of nonpolarizable systems in the
infinite wavevector and frequency limit, $\varepsilon_\infty=1$, can only be
obtained if the dielectric constant is known up to about $k \sim 150 k_{\rm
min} \approx 50 {\rm \AA}^{-1}$ and $\omega \sim 1000$ THz. It is worth to
remark also that at $k \ge 20 k_{\rm min}$ the frequency dependence of the
dielectric constant can be reproduced quantitatively using the analytical
formula (A3) for the time correlation functions of a noninteracting system.
The dielectric constant, calculated in such a way, is shown in fig.~6f-h
and fig.~7 by circles.

The dielectric constant $\varepsilon \scs{\rm L}(k,\omega)$ at fixed real
values of wavevector and frequency describes the response of the system on
longitudinal electric fields in the form of monochromatic plane waves, i.e.,
when $E(\bms{r},t) \sim {\mbox{\large e}}^{{\rm i}(\omega t-\bvs{k\!\cdot
\!r})}$. Arbitrary inhomogeneous fields can be presented by a set of the
plane waves via the time and spatial Fourier transform. Therefore, the
response on such fields can be determined integrating the corresponding
monochromatic contributions over the frequency and wavenumber spectrum
using $\varepsilon \scs{\rm L}(k,\omega)$. The necessity of introducing
pure imaginary frequencies $\omega \equiv {\rm i} \omega^*$ with $\omega^*>0$
arises in the problem of finding electric fields $E(\bms{r},t) \sim {\mbox
{\large e}}^{(-\omega^* t -{\rm i}\bvs{k\!\cdot\!r})}$ in the system when
the external field is turned off. Such damping in time electric fields can
be obtained using the dispersion relation $\varepsilon \scs{\rm L}(k,\omega)
=0$ [28]. It can be seen from figs.~6, 7 that this relation is not satisfied
at real frequencies. So that, contrary to transverse electromagnetic waves,
longitudinal monochromatic waves can not propagate without external sources
of charges. From the fluctuation formula (4) it follows that $\varepsilon
\scs{\rm L}(k,\omega) \to 0$ when ${\scr L}_{{\rm i}\omega} (-\dot g
\scs{\rm L}(k,t)) \to \infty$. The last limit can be achieved at imaginary
frequencies $\omega = {\rm i} \omega^*$ only, where $\omega^*(k) \to
1/\tau_{\rm L}^{\rm rel}(k)$. Thus, the electric fields are damped
exponentially with the characteristic interval of order of the relaxation
time. Maximal life times of the longitudinal electric excitations are
observed at spatial inhomogeneity of $k \sim 1.7 {\rm \AA}^{-1}$ (see
fig.~5) and consists about 1.6ps. Strong spatially inhomogeneous fields
($k \to \infty$) disappear immediately after disappearing external
supporting fields.

Finally, we consider some aspects of applying the well-known Kramers-Kronig
relations to the longitudinal wavevector- and frequency-dependent dielectric
constant. In general, these relations connect real $\xi'(\omega)$ and
imaginary $\xi''(\omega)$ parts of the function $\xi(\omega) = \xi'(\omega)-
{\rm i} \xi''(\omega)$ of frequency as
\begin{equation}
\xi'(\omega)=\frac2\pi \int \limits_0^\infty
\hspace{-9.32pt}\mbox{\large \bf - }
\frac{x \xi''(x)}{x^2-\omega^2} {\rm d} x \ , \ \ \ \ \ \ \ \
\xi''(\omega)=-\frac{2\omega}\pi \int \limits_0^\infty
\hspace{-9.32pt}\mbox{\large \bf - }
\frac{\xi'(x)}{x^2-\omega^2} {\rm d} x \ .
\end{equation}
From the mathematical point of view the relations (15) are fulfilled if the
function $\xi(\omega)$ can be presented in the form ${\scr L}_{{\rm i}\omega}
(\xi(t))$, where $\xi(t)$ is a real analytical function of time. Then, taking
into account the fluctuation formulas (4), it can be shown that in two cases
at least, namely, when $\xi(\omega) \equiv \varepsilon \scs{\rm L}(0,\omega)
-1=\varepsilon(\omega)-1$ and $\xi(\omega) \equiv \Big(\varepsilon \scs{\rm
L}(k,\omega)-1\Big) \Big/\varepsilon \scs{\rm L}(k,\omega)$ for arbitrary
$k$, the Kramers-Kronig relations will be satisfied. At the same time it is
not obvious that these relations must be valid for the function $\xi(\omega)
\equiv \varepsilon \scs{\rm L}(k,\omega)-1$ at $k \ne 0$. We have established
on the basis of our numerical analysis that the Kramers-Kronig relations can
apply to the longitudinal dielectric constant in the form $\varepsilon
\scs{\rm L}(k,\omega)-1$ at such wavenumbers only, where the static
dielectric constant $\varepsilon \scs{\rm L}(k)$ takes positive values,
i.e., when $0 \le k < k \scs{\rm I}$ and $k > k \scs{\rm II}$. Physically,
the impossibility to use the function $\varepsilon \scs{\rm L}(k,\omega)-1$
for the transformations (15) at arbitrary wavenumbers lies in the fact that,
in fact, the external field $\bms{E} \scs{0}$, but not the total field
$\bms{E}$, can be considered as an independent parameter which causes
time varying all observing quantities in the system.

\vspace{12pt}

\section{Conclusion}

\hspace{1em}  A fluctuation formula suitable for the self-consistent
calculations of the wavevector- and frequency-dependent dielectric constant
for interaction site models of polar systems has been rigorously derived in
the situation that is typical for computer simulations. Using this formula,
the longitudinal component of the dielectric constant has been evaluated
by the MD method for the TIP4P model of water in a very wide scale of
wavenumbers and frequencies. The most striking feature of interaction site
models consists in existing of the libration oscillations in the shape of
longitudinal time-dependent polarization fluctuations and this feature is
reproduced by the TIP4P potential as well. We have showed, however, that
the libration oscillations vanish at increasing wavenumber. Choosing the
correct microscopic variable for polarization density, which corresponds
completely to interaction site models, allows us to investigate the
frequency dependence of the dielectric constant for arbitrary wavevector
values. At the same time, it has been corroborated by the explicit
calculations that the PD approximation is unsuitable, in general, for
evaluating the wavevector- and frequency-dependent dielectric constant
of interaction site models of polar fluids.

Since it has now been shown that the calculation of the wavevector- and
frequency-dependent dielectric constant in computer simulations for a given
interaction site model is practical in principle, we believe that this fact
will stimulate further research in both theory and pure experiment. A next
paper of this series will be devoted to a more complicated case, namely, to
the calculation of the transverse wavevector- and frequency-dependent
dielectric constant for the TIP4P water.

\vspace{24pt}

The author would like to acknowledge financial support of the President
of Ukraine.

\newpage
\small

\begin{center}
{\Large \bf Appendix}
\end{center}
\setcounter{equation}{0}
\renewcommand{\theequation}{A\arabic{equation}}

We now consider the time dependence of the longitudinal wavevector-dependent
Kirkwood factor
\begin{equation}
g \scs{\rm L}(k,t) = \frac{1}{N \mu^2 k^2} \left< \sum_{i,j; a,b}^{N;M}
q \scs{a} q \scs{b} {\mbox{\large e}}^{{\rm i} \bvs{k} \bvs{\cdot}
(\bvs{r}_i^a(t)-\bvs{r}_j^b(0))} \right>_0
\end{equation}
in the free motion regime. In this case, molecules are statistically
independent of one another and, therefore, nonzero contributions to
$g \scs{\rm L}(k,t)$ give only the self (intramolecular) part of terms
with coincident molecular indexes $(i=j)$ of summation (A1). The site
velocities of molecules can be presented as $\bms{v}_i^a(t)=\bms{V\!}_i+
\bms{\mit \Omega}_i(t) \bms{\times} \bms{\delta}_i^a(t)$, where $\bms{V\!}_i$
and $\bms{\mit \Omega}_i$ are the translational and angular velocities of
the $i$th molecule, respectively. We note that according to the Euler
equations, the angular velocities depend on time even for free rotational
motion, so that $\bms{r}_i^a(t)=\bms{r}_i(0)+\bms{V\!}_i\,t+\bms{\delta}_i^a
(t)$, where $\bms{\delta}_i^a(t)=\bms{\delta}_i^a(0)+\int_0^t [\bms{\mit
\Omega}_i(t') \bms{\times} \bms{\delta}_i^a(t')] {\rm d} t'$. Then, taking
into account that the translational and angular velocities of each molecule
are distributed independently at equilibrium, the expression (A1) transforms
into
\begin{equation}
g^{\rm (f)} \scs{\rm L}(k,t) = \frac{1}{N \mu^2 k^2} \sum_{i=1}^N \left<
{\mbox{\large e}}^{{\rm i} \bvs{k} \bvs{\cdot} \bvs{V}_i t} \right>_0
\left< \sum_{a,b}^{M} q \scs{a} q \scs{b}
{\mbox{\large e}}^{{\rm i} \bvs{k}
\bvs{\cdot} (\bvs{\delta}_i^a(t)-\bvs{\delta}_i^b(0))}
\right>_0 .
\end{equation}

The first averaging in (A2) can easily be evaluated applying the Maxwell
distribution $f \scs{0} (V_\alpha)= \left( \frac{m}{2\pi k_{\rm B} T}
\right)^{1/2} \exp \left(-\frac{m V_\alpha^2}{2  k_{\rm B} T} \right)$
with respect to the cartesian components ($\alpha=x,y,z$) of the
translational velocity $\bms{V}$, where $m$ is the mass of the molecule.
As a result, we obtain $\left< {\mbox{\large e}}^{{\rm i} \bvs{k} \bvs
{\cdot} \bvs{V} t} \right>_0=\displaystyle \int\!\int\!\int_{-\infty}^
\infty {\rm d} V_x {\rm d} V_y {\rm d} V_z f \scs{0} (V_x) f \scs{0} (V_y)
f \scs{0} (V_z) \exp({\rm i}(k_x V_x+k_y V_y+k_z V_z))=\textstyle \exp
\left(- \frac{k_{\rm B} T}{2m} k^2 t^2 \right)=f \scs{\rm G}(k,t)$. The
second averaging in (A2) is not reduced, in general, to an analytic form.
However, due to the fact that the Gaussian multiplier $f \scs{\rm G}(k,t)$
quickly decays to zero with increasing $t$ at $k \ne 0$, we can cast the
solution for $\bms{\delta}_i^a(t)$ into the Taylor series by writing
$\bms{\delta}_i^a(t) = \bms{\delta}_i^a + [\bms{\mit \Omega}_i \bms{\times}
\bms{\delta}_i^a] t + {\cal O}(t^2)$, where $\bms{\mit \Omega}_i$ and
$\bms{\delta}_i^a$ are taken at time $t=0$ and terms of order $t^2$ and
higher have been neglected. The integration over angular velocities it is
convenient to perform in the moving coordinate system $XYZ$, attached to
the molecule, in which the Maxwell distribution has the form $f \scs{0}
(\Omega_\alpha)= \left( \frac{J_\alpha}{2\pi k_{\rm B} T} \right)^{1/2}
\exp \left(-\frac{J_\alpha \Omega_\alpha^2}{2  k_{\rm B} T} \right)$,
where $J_\alpha$ denote the moments of inertia along principal axes
$\alpha=X,Y,Z$. Finally, performing averaging over orientations of the
molecule with respect to the laboratory frame $xyz$, we obtain the
desired result
\begin{eqnarray}
g^{\rm (f)} \scs{\rm L}(k,t) \!\!\!\!\!\!\!\!\!&&=\frac{f \scs{\rm G}(k,t)}
{4\pi \mu^2 k^2} \sum_{a,b}^{M} q \scs{a} q \scs{b} \int_0^\pi \sin \theta
{\rm d} \theta \int_0^{2\pi} {\rm d} \phi \cos(k d_{ab} \cos \theta)
\exp \left( -\frac{k_{\rm B}T}{2} k^2 t^2  \right. \nonumber \\ [2pt] &&
\left. \times \left\{ \frac{(\Delta_Y^a \cos \theta-\Delta_Z^a \sin \theta
\sin \phi)^2}{J_X} + \frac{(\Delta_X^a \cos \theta-\Delta_Z^a \sin \theta
\cos \phi)^2}{J_Y} \right. \right. \\ [2pt] && \left. \left. +
\frac{(\Delta_X^a \sin \theta \sin \phi-\Delta_Y^a \sin \theta
\cos \phi)^2}{J_Z} \right\} \right) + {\cal O}\left( \frac{t^2}{k^2}
\right) , \ \ \ \nonumber
\end{eqnarray}
where $(\Delta_X^a, \Delta_Y^a, \Delta_Z^a)$ designate the principal
components of $\bms{\delta}_i^a$ and $d_{ab}=|\bms{\delta}_i^a-
\bms{\delta}_i^b|$ are the distances between charges $a$ and $b$
within the same molecule.

At a given molecular geometry the two-dimensional integral (A3) can
easily be calculated numerically. In the static limit $(t=0)$ this
integral is taken analytically,
\begin{equation}
g^{\rm (f)} \scs{\rm L}(k) = \frac{1}{\mu^2 k^2} \left(
\sum_{a=1}^M q^2 \scs{a}+\sum_{a \ne b}^M q \scs{a} q \scs{b}
\frac{\sin(k d_{ab})}{k d_{ab}} \right) .
\end{equation}
It is worth to remark that the formulas (A3), (A4) at sufficiently large
values of wavevector $k$ can be applied to interacting systems as well,
because then both intermolecular terms in (A1) and terms in (A3)
caused by interactions (nonlinear in $t$) are small.

\newpage
\renewcommand{\thepage}{}

\begin{center}
{\large Figure captions}
\end{center}

\addtolength{\baselineskip}{-1pt}

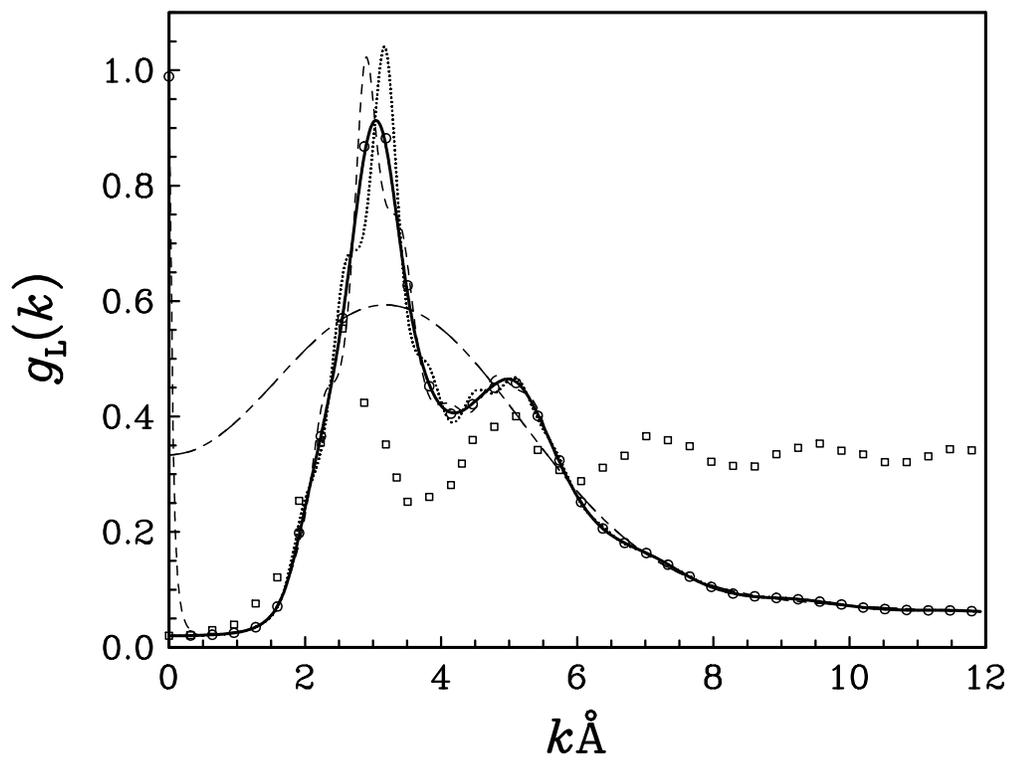
{\bf Fig.~1.}~The longitudinal wavevector-dependent Kirkwood factor of
the TIP4P water. The result obtained for the finite and infinite systems
within the IS description using the ISRF geometry is plotted by dashed and
solid curves, respectively. The corresponding result of the point dipole
approximation for the infinite system is presented by open squares. The
infinite-system Kirkwood factor, evaluated in the Ewald and PDRF geometries,
is shown as open circles and dotted curve, respectively. The long-short
dashed curve corresponds to the Kirkwood factor in the free motion regime
(equation (A4)).

\vspace{12pt}

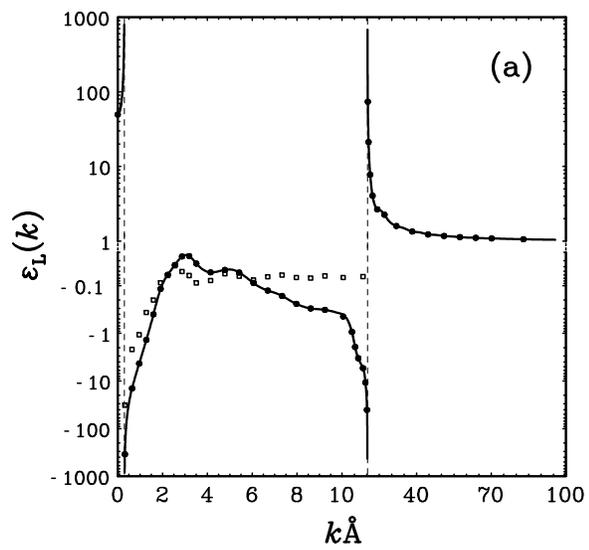
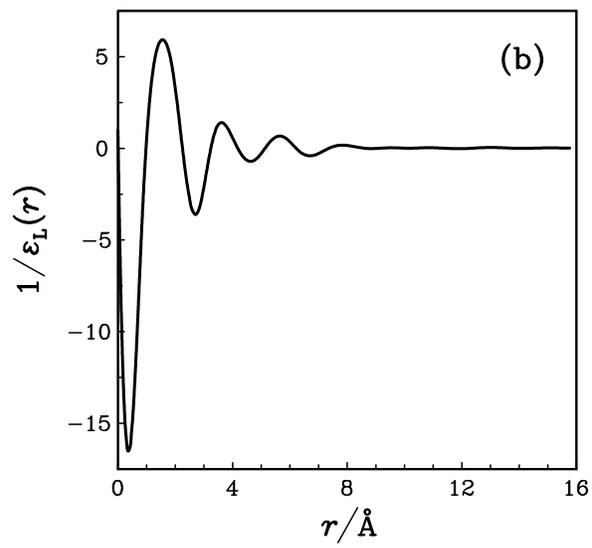
{\bf Fig.~2.}~The longitudinal wavevector-dependent dielectric constant
{\bf (a)} of the TIP4P water within the IS description (circles connected by
the solid curve) and the PD approximation (open squares). Two vertical lines
indicate the positions of singularities. The function, describing the static
screening of external charges in water, is shown in {\bf (b)}.

\vspace{12pt}

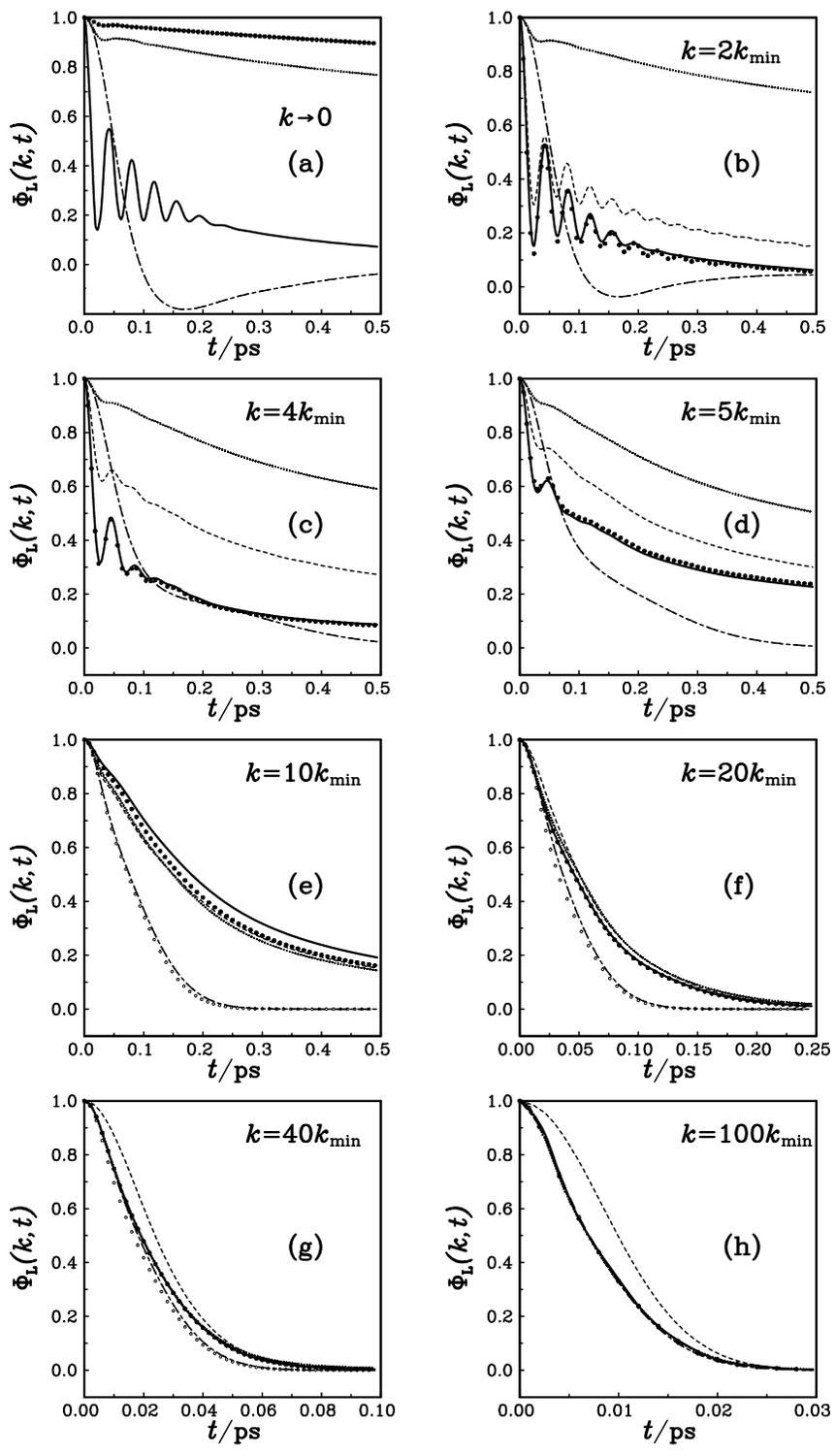
{\bf Fig.~3.} The normalized, time autocorrelation functions of the
longitudinal polarization fluctuations in the TIP4P water for the finite
(circles) and infinite (solid curves) systems in the IS description as well
as for the finite system in the PD approximation (dashed curves). The dotted
curves show the self parts of the IS functions. The correlation functions of
a noninteracting system, calculated using the exact relation (A2) and the
short time approximation (A3), are shown by the long-short dashed curves and
open circles, respectively. We note that five of the six dependencies
are undistinguished in {\bf (h)}.

\vspace{12pt}

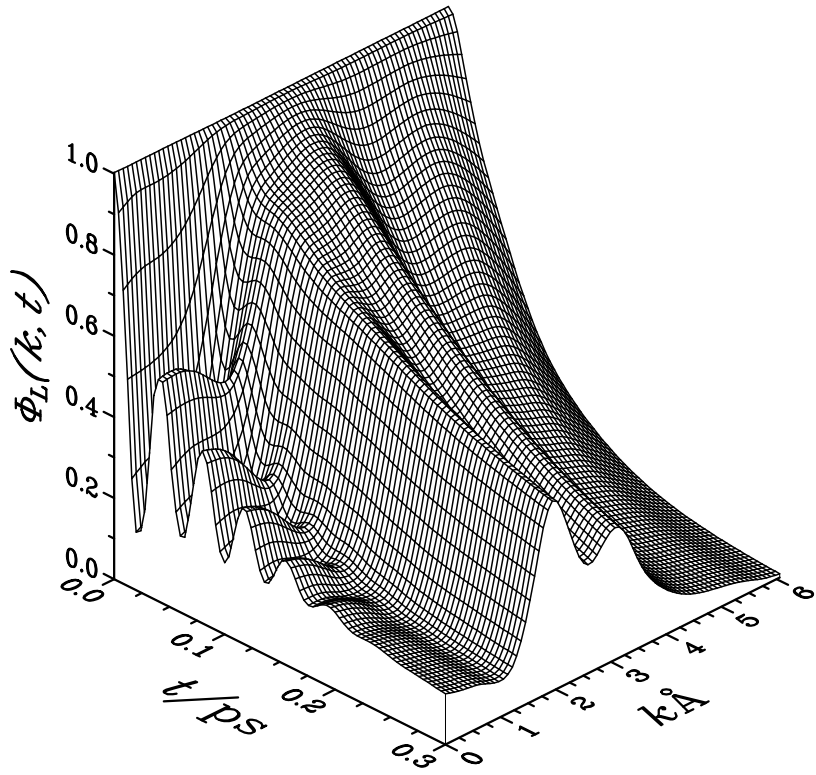
{\bf Fig.~4.} The 3D plot of the wavevector- and time-dependent longitudinal
polarization fluctuations in the IS description for the infinite system.

\vspace{12pt}

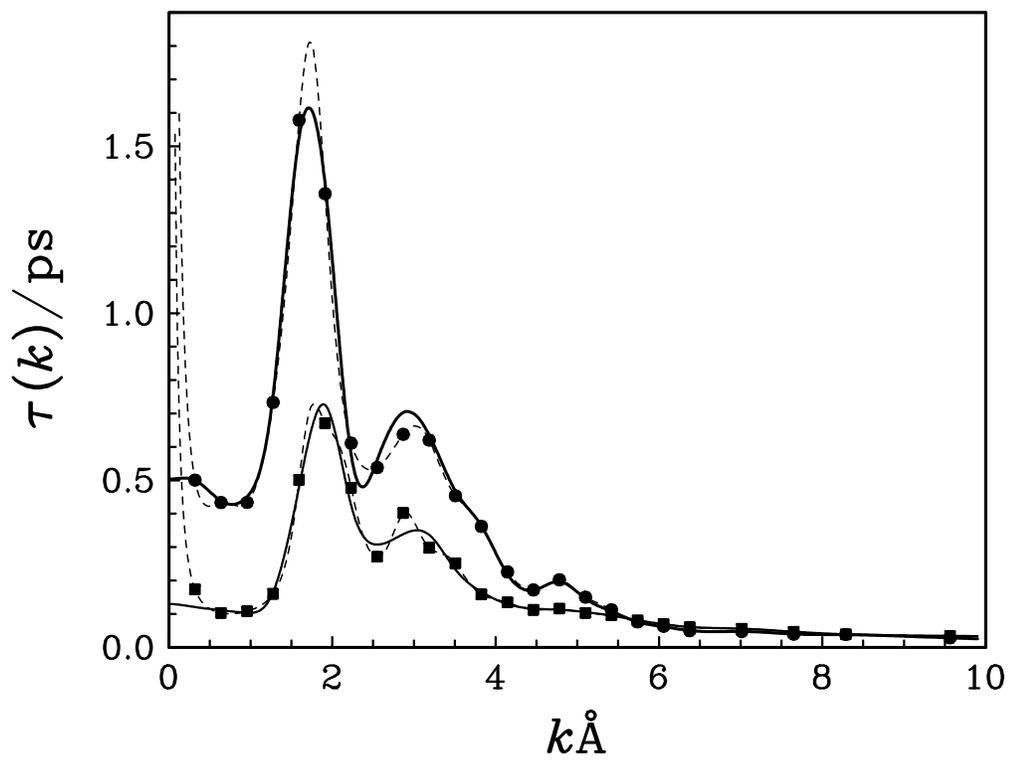
{\bf Fig.~5.} The wavevector-dependent relaxation (circles) and correlation
(squares) times for the longitudinal polarization fluctuations of the
finite system. The results for the infinite systems are shown by the
corresponding solid curves.

\vspace{12pt}

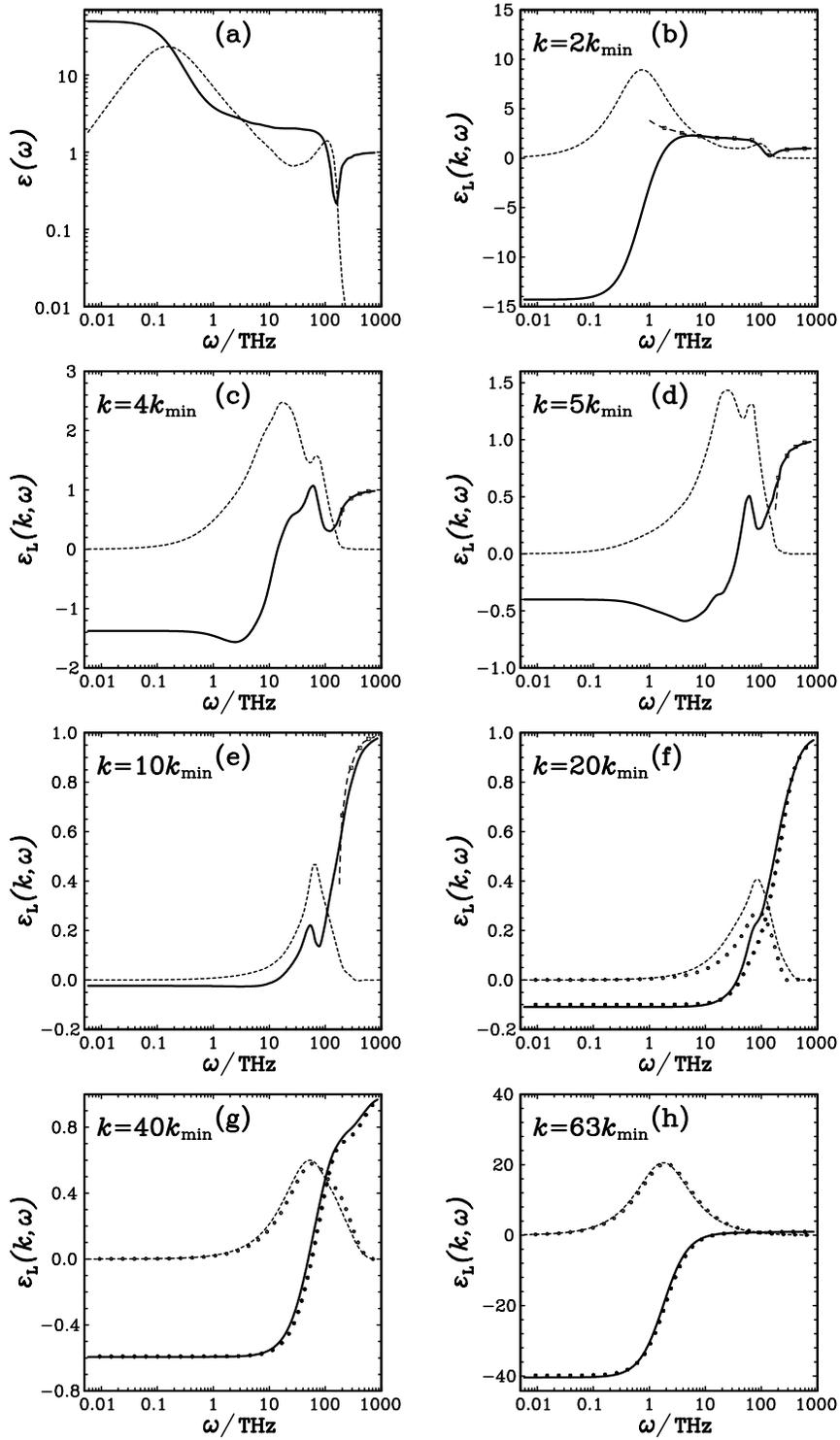
{\bf Fig.~6.} The longitudinal wavevector- and frequency-dependent
dielectric constant of the TIP4P water. The real and imaginary parts
are shown as solid and dashed curves, respectively. Note the logarithmic
scale in {\bf (a)}. Open squares connected by the long dashed curve in
{\bf (b)}-{\bf (e)} reproduce the real part of the frequency-dependent
dielectric constant at zero wavevector. The results obtained in the
free motion regime (equation (A3)) are plotted in {\bf (f)}-{\bf (h)}
as full (real part) and open (imaginary part) circles, respectively.

\vspace{12pt}

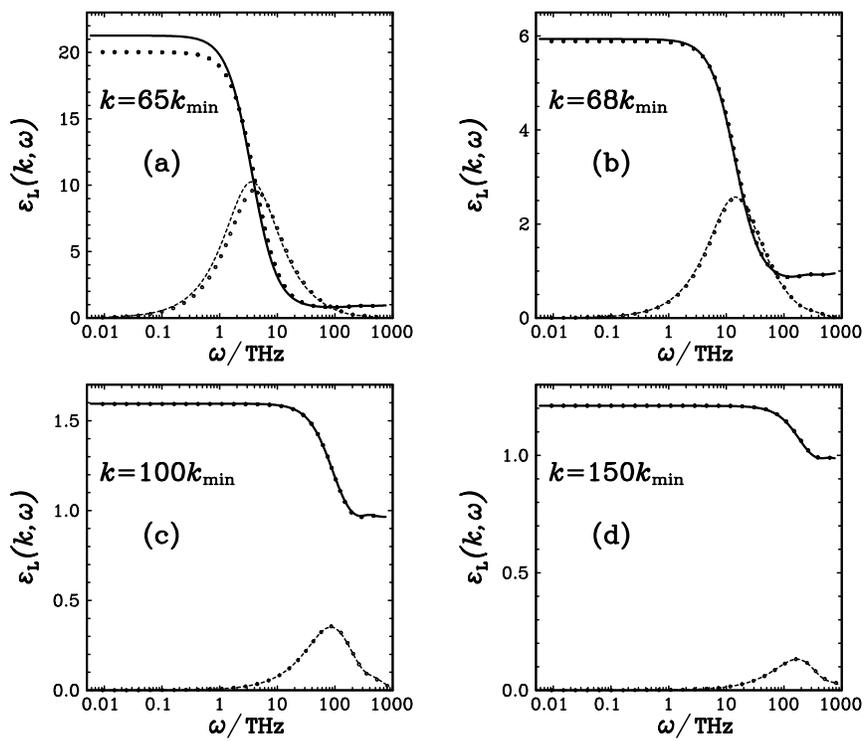
{\bf Fig.~7.} The longitudinal wavevector- and frequency-dependent
dielectric constant of the TIP4P water at great wavenumber values.
Other features as for fig.~6.

\end{document}